\documentclass[11pt]{article}
\usepackage{amsmath}
\usepackage{amssymb}
\begin{document}
\baselineskip=17pt
\begin{titlepage}
\begin{flushright}
{\small TUM-HEP-621/06}\\[-1mm]
{\small KYUSHU-HET-93}\\[-1.5mm]
hep-ph/0603116
\end{flushright}
\begin{center}
\vspace*{8mm}

{\large\bf Twisted Flavors and Tri/bi-Maximal Neutrino Mixing}
\vspace*{8mm}

Naoyuki Haba$^{\rm a},\,$ Atsushi Watanabe$^{\rm b},\,$ and 
Koichi Yoshioka$^{\rm b}$
\vspace*{3mm}

{\it $^{\rm a}$Physik-Department, Technische Universit\"at 
M\"unchen,}\\[-1mm]
{\it James-Franck-Stra{\ss}e, 85748 Garching, Germany}\\
{\it $^{\rm b}$Department of Physics, Kyushu University, Fukuoka,
812-8581, Japan}
\vspace*{2mm}

{\small (March, 2006)}
\end{center}
\vspace*{4mm}

\begin{abstract}\noindent%
A new framework for handling flavor symmetry breaking in the neutrino
sector is discussed where the source of symmetry breaking is traced to
the global property of right-handed neutrinos in extra-dimensional
space. Light neutrino phenomenology has rich and robust predictions
such as the tri/bi-maximal form of generation mixing, controlled mass
spectrum, and no need of flavor mixing couplings in the theory.
\end{abstract}
\end{titlepage}

The fermion masses and mixing angles have been long-standing as well
as inspiring problems in particle physics in these decades. Their
observed patterns, particularly the recent results of neutrino
experiments, strongly suggest profound principles such as grand
unification and flavor symmetry lying behind the data. However flavor
symmetry should be broken in non-trivial way while preserving the trace
of symmetry structure. In this letter we present a novel way to
realize flavor symmetry breaking in the lepton sector where a central
ingredient is the global property of right-handed neutrinos in
extra-dimensional space. Our method provides highly predictive,
calculable effects of flavor breaking, and is also applied to the
quark sector and grand unified theory.

\medskip
As a simple example of higher-dimensional theory, we consider a
five-dimensional model where the standard-model fields including
three-generation lepton doublets $l_i$ ($i=1,2,3$) are confined on the
four-dimensional wall at $x^5=0$. Besides the gravity, only
standard-model gauge singlet fields can propagate in the
bulk~\cite{neuExD}\cite{neuExD2} not to violate the charge
conservations. We thus introduce gauge singlet five-dimensional Dirac
fermions $\Psi_i$ ($i=1,2,3$) which mean right-handed 
neutrinos $\psi_{R_i}$ and their chiral partners. The Lorentz
invariance leads to the general five-dimensional Lagrangian
\begin{eqnarray}
  {\cal L} &=& i\overline{\Psi_j}\partial\!\!\!/\Psi_j
  -\overline{\Psi_i}\,m_{D_{ij}}\Psi_j 
  -\frac{1}{2}\big(\overline{\Psi^c_i}\,M_{ij}\,\Psi_j\,
  +\text{h.c.}\big)  \nonumber \\[1mm]
  && \qquad -\big(\overline{l_i}\,m_{ij}\Psi_j+
  \overline{l_i}\,m^c_{ij}\Psi^c_j\big)\,\delta(x^5) \,+{\rm h.c.},
\end{eqnarray}
where $m$, $m^c$ denote the mass matrices after the electroweak
symmetry breaking, and the conjugated spinor is defined 
by $\Psi^c=\Gamma^3\Gamma^1\overline\Psi{}^t$ so that it is Lorentz
covariant in five dimensions. The existence of four-dimensional
subspace violates the bulk Lorentz invariance locally. If the fifth
component of a vector field develops a nonzero expectation value, the
five-dimensional Lorentz invariance is broken in the whole bulk and
that could induce mass terms $\overline\Psi\Gamma^5m_D^5\Psi$, 
$\overline{\Psi^c}\Gamma^5M^5\Psi$~\cite{LRRR}, but we do not consider
such a possibility in this letter.

\medskip
Let us suppose the extra space is compactified on 
the $S^1/Z_2$ orbifold with radius $R$. There are two types of
operations on this space; the reflection $\hat Z$~:~$x^5\to-x^5$ and
the translation $\hat T$~:~$x^5\to x^5+2\pi R$. They trivially 
satisfy $\hat T\hat Z=\hat Z\hat T^{-1}$. On field variables, these
operations are expressed in terms of matrices in the field 
space: $\hat Z\phi(x^5)=Z^{-1}\phi(-x^5)$ and $\hat T\phi(x^5)=
T^{-1}\phi(x^5+2\pi R)$. The boundary conditions are then defined by
the identifications $\hat Z\phi(x^5)=\phi(x^5)$ and $\hat T\phi(x^5)=
\phi(x^5)$. In particular, for the bulk fermions $\Psi_i$, they are
explicitly given by
\begin{equation}
  Z \,=\, X_Z\otimes i\Gamma^5, \qquad 
  T \,=\, X_T\otimes 1,
\end{equation}
where $X_Z$ and $X_T$ are $3\times3$ unitary matrices acting on the
generation space, which satisfy the consistency
relation $X_ZX_T=X_T^{-1}X_Z$. The $i\Gamma^5$ part is needed to make
the kinetic term invariant, and it also turns out to forbid the bulk
Dirac mass $m_D$. Exactly speaking, the parity $Z$ has an ambiguity
of sign, but it has an effect that exchanges upper and lower
components of a Dirac fermion and is physically irrelevant to later
discussion.

With a set of boundary conditions, the bulk fermions $\Psi_i$ are
expanded by Kaluza-Klein modes with their kinetic terms properly
normalized: 
\begin{equation}
\Psi_i(x^\mu,x^5)\;=\;
\Big(\sum_m\big[W_R^{(m)}(x^5)\big]_{ij}\psi_{R_j}^{(m)}(x^\mu),\;
\sum_n\big[W_L^{(n)}(x^5)\big]_{ij}\psi_{L_j}^{(n)}(x^\mu)\,\Big)^{\rm t}.
\end{equation}
After integrating over the fifth dimension, we obtain the neutrino
mass matrix in four-dimensional effective theory:
\begin{equation}
    {\cal L}_{\rm mass} \;=\;
  -\frac{1}{2}{\cal N}^{\rm t}\epsilon {\cal M}{\cal N} 
  \,+{\rm h.c.},
\end{equation}
\begin{equation}
\renewcommand{\arraystretch}{1.6}
{\cal M} \;=\; \left(%
\begin{array}{c|cc|cc|c}
& -M_0 & M^c_0 & -M_1 & M^c_1 & \cdots \\ \cline{1-6}
\!-M_0^{\rm t} & -M_{R_{00}} & M_{D_{00}}^{\rm t} & -M_{R_{01}}
& M_{D_{10}}^{\rm t} & \cdots \\
\,M^c_0{}^{\rm t} & \phantom{-}M_{D_{00}} & M_{L_{00}}^* & 
\phantom{-}M_{D_{01}} & M_{L_{01}}^* & \cdots \\ \cline{1-6}
\!-M_1^{\rm t} & -M_{R_{10}} & M_{D_{01}}^{\rm t} & -M_{R_{11}}
& M_{D_{11}}^{\rm t} & \cdots \\
\,M^c_1{}^{\rm t} & \phantom{-}M_{D_{10}} & M_{L_{10}}^* & 
\phantom{-}M_{D_{11}} & M_{L_{11}}^* & \cdots \\ \cline{1-6}
\vdots & \vdots & \vdots & \vdots & \vdots & \ddots
\end{array}\right),
\end{equation}
where
\begin{eqnarray}
  M_{D_{mn}} \!&=& 
  \int\!dx^5\; W_L^{(m)}{}^\dagger\partial_y W_R^{(n)}, \\[1mm]
  M_{R_{mn}} \!&=&
  \int\!dx^5\; W_R^{(m)}{}^{\rm t} M\,W_R^{(n)}\,, \\[1mm]
  M_{L_{mn}} \!&=&
  \int\!dx^5\; W_L^{(m)}{}^{\rm t} M\,W_L^{(n)}\,, \\[2mm]
  M_n \;&=& m W_R^{(n)}(0), \\[2mm]
  M^c_n \;&=& m^c W_L^{(n)}(0)^*\,.
\end{eqnarray}
The mass matrix ${\cal M}$ is written in the 
basis ${\cal N}=(\epsilon l_i^*,\,\psi_{R_i}^{(0)},\,
\epsilon\psi_{L_i}^{(0)*},\,\psi_{R_i}^{(1)},\,
\epsilon\psi_{L_i}^{(1)*},\cdots)\,$ where zero modes are suitably
extracted according to boundary conditions ($\epsilon=i\sigma^2$).

\medskip
In this letter we consider $S_3$ as an example of flavor symmetry
group, which is the simplest non-abelian discrete group and has
attractive features as the flavor symmetry~\cite{S3}. The $S_3$ group
is composed of six elements which are the identity $X_1=I$, two types
of cyclic rotations $X_{2,3}$, and three types of 
permutations $X_{4,5,6}$. The non-trivial representations are 
doublet $\underline{2}$ and pseudo singlet $\underline{1}_P$, the
latter of which changes the sign under a permutation. On 
the $\underline{3}=\underline{2}+\underline{1}$ representation, the
group elements are expressed by the following $3\times3$ matrices
\begin{eqnarray}
  &&
  X_1 =\begin{pmatrix} 1 & & \\ & 1 & \\ & & 1 \end{pmatrix},\quad
  X_2 =\begin{pmatrix} & & 1 \\ 1 & & \\ & 1 & \end{pmatrix},\quad
  X_3 =\begin{pmatrix} & 1 & \\ & & 1 \\ 1 & & \end{pmatrix}, 
  \nonumber \\[1mm]
  &&
  X_4 =\begin{pmatrix} 1 & & \\ & & 1 \\ & 1 & \end{pmatrix},\quad
  X_5 =\begin{pmatrix} & & 1 \\ & 1 & \\ 1 & & \end{pmatrix},\quad
  X_6 =\begin{pmatrix} & 1 & \\ 1 & & \\ & & 1 \end{pmatrix}.
\end{eqnarray}
It is interesting to note that possible boundary conditions for bulk
fields are limited by several consistencies. First, since $\hat Z$ is 
a parity, available choices are $X_Z=X_{1,4,5,6}$. In addition, for
the translation, $X_T$ must satisfy the 
relation $X_ZX_T=X_T^{-1}X_Z$. Clearly not all types of boundary
conditions are allowed; even in the case that $X_Z$ is 
trivial, $X_T$ is limited to $X_{1,4,5,6}$. That is more clearly seen
by defining another parity operation $\hat Z'=\hat T\hat Z$, which is
the reflection about $x^5=\pi R$.

\medskip
Now let us discuss flavor symmetry breaking with boundary twisting and
characteristic consequences on neutrino physics. Throughout this
letter, as an example, the three generations of lepton doublets and
bulk fermions are set to be $S_3$ triplets in which the first and
second generations constitute doublets and the third ones are
singlets. The first case we consider is along the line of the original
Scherk-Schwarz theory~\cite{SS}, that is, to have twisted flavors
around the extra dimension: ($X_Z,X_T)=(1,X_4$) on the bulk 
fermions $\Psi_i$. With this condition at hand, the mode expansion
becomes ($n\geq1$)
{\allowdisplaybreaks%
\begin{eqnarray}
W_R^{(0)} \!&=& \frac{1}{\sqrt{2\pi R}}\begin{pmatrix}
  1 & ~~ & \\[1mm]
  & & \frac{1}{\sqrt{2}} \\[2mm]
  & & \frac{1}{\sqrt{2}}
\end{pmatrix}, \\[2mm]
W_R^{(n)} \!&=& \frac{1}{\sqrt{\pi R}}\begin{pmatrix}
  \cos\big(n\frac{x^5}{R}\big) & & \\[2mm]
  & \!\frac{-1}{\sqrt{2}}\cos\big[\big(n-\frac{1}{2}\big)
  \frac{x^5}{R}\big] 
  & \frac{1}{\sqrt{2}}\cos\big(n\frac{x^5}{\!R}\big) \\[2mm]
  & \!\frac{1}{\sqrt{2}}\cos\big[\big(n-\frac{1}{2}\big)
  \frac{x^5}{R}\big] 
  & \frac{1}{\sqrt{2}}\cos\big(n\frac{x^5}{R}\big)
\end{pmatrix}, \\[2mm]
W_L^{(n)} \!&=& \frac{1}{\sqrt{\pi R}}\begin{pmatrix}
  \sin\big(n\frac{x^5}{R}\big) & & \\[2mm]
  & \!\frac{-1}{\sqrt{2}}\sin\big[\big(n-\frac{1}{2}\big)
  \frac{x^5}{R}\big] 
  & \frac{1}{\sqrt{2}}\sin\big(n\frac{x^5}{\!R}\big) \\[2mm]
  & \!\frac{1}{\sqrt{2}}\sin\big[\big(n-\frac{1}{2}\big)
  \frac{x^5}{R}\big]
  & \frac{1}{\sqrt{2}}\sin\big(n\frac{x^5}{R}\big)
\end{pmatrix}.
\end{eqnarray}}%
The mass terms in the Lagrangian are generally described by two types
of flavor symmetric forms: the unit matrix $I$ and the democratic 
matrix $D$ where all elements are equal to 1;
\begin{equation}
  M \;=\; aI+\frac{b}{3}D, \qquad\quad m \;=\; pI+\frac{r}{3}D.
\end{equation}
Another mass term $m^c$ is irrelevant in the present case 
since $X_Z=1$ and the lower components decouple from the
standard-model fields [$W_L^{(n)}(0)=0$]. Integrated out the infinite
tower of bulk neutrinos, the standard-model neutrinos $l_i$ receive
the Majorana masses
\begin{equation}
  {\cal M}_l \;=\;
  \frac{\frac{|a+b|}{a+b}\,(p+r)^2}{6\tanh(|a+b|\pi R)}\,D
  +\frac{\frac{|a|}{a}p^2}{12\tanh(|a|\pi R)}\,E 
  +\frac{\frac{|a|}{a}p^2\tanh(|a|\pi R)}{4}\,F,
\end{equation}
in a good approximation that 
$\frac{|p|}{\sqrt{R}},\,\frac{|r|}{\sqrt{R}}\ll |a|,|b|$ 
or $1/R$, i.e.\ the Dirac masses are much smaller than the bulk
Majorana masses or the compactification scale. We have defined the
integer matrices
\begin{equation}
 E = \begin{pmatrix}
   4 & -2 & -2 \\[1mm] -2 & 1 & 1 \\[1mm] -2 & 1 & 1
 \end{pmatrix}, \qquad
 F = \begin{pmatrix}
   ~~~ & & \\[1mm] & 1 & -1 \\[1mm] & -1 & 1
 \end{pmatrix}.
\end{equation}
The obtained matrix ${\cal M}_l$ gives notable predictions of neutrino
masses and mixing angles. The mass eigenvalues are found to be
\begin{eqnarray}
  |m_1| &=& \frac{|p|^2}{2\tanh(|a|\pi R)}, \nonumber \\[.5mm]
  |m_2| &=& \frac{|p+r|^2}{2\tanh(|a+b|\pi R)}, \nonumber \\[1.5mm]
  |m_3| &=& \frac{|p|^2\tanh(|a|\pi R)}{2}. 
\end{eqnarray}
The current experimental status proves possible three types of mass
spectra~\cite{review}; normal hierarchy (NH), inverted hierarchy (IH),
and degenerate neutrinos. It is found that the latter two spectra are
accommodated to the above mass predictions. (If a Lorentz-violating
bulk Majorana mass were adopted instead of $M$, $\tanh$ is replaced
with $\tan$ in the mass formula, and NH can be obtained.) \ For
example, the IH case is realized for a small bulk Majorana 
mass ($|a|R\ll1$) and suppressed democratic couplings $b$ and $r$.
Typical mass scales are $|m_{1,2}|\sim10^{-1}\,\text{eV}\,
\big(\frac{10^8}{|a|\;\text{[GeV]}}\big)
\big(\frac{10^{-10}}{R\;\text{[GeV$^{-1}$]}}\big)^\frac{2}{3}$ and
$|m_3|\sim10^{-4}\,\text{eV}\,
\big(\frac{|a|\;\text{[GeV]}}{10^8}\big)
\big(\frac{R\;\text{[GeV$^{-1}$]}}{10^{-10}}\big)^\frac{4}{3}$.
Similar estimations can also be done for other patterns. In any case,
one has almost no need of generation mixing couplings both in the
Dirac and bulk Majorana mass terms in order for the mass eigenvalues
being acceptable. The mass hierarchy structure is also interesting. In
usual case, the smaller heavy-field Majorana masses, the larger
induced light neutrino masses. However we have found in the above the 
eigenvalue $m_3$ has an opposite behavior: a smaller bulk Majorana
mass ($|a|R\ll1$) rather suppresses induced mass ($m_3$) than
usual. This is because such eigenvalues come only from heavy excited
modes which are decoupled in the limit $|a|R\ll1$. Such unconventional
behavior is useful not only for generating tiny neutrino masses but
also to discuss some neutrino physics in cosmology. Contrary to wide
possibility of spectrum, the neutrino mixing matrix is definite and
given by
\begin{equation}
V \;=\;
\begin{pmatrix}
  \frac{2}{\sqrt{6}} & \frac{1}{\sqrt{3}} & 0 \\[2mm]
  \frac{-1}{\sqrt{6}} & \frac{1}{\sqrt{3}} & \frac{-1}{\sqrt{2}} \\[2mm]
  \frac{-1}{\sqrt{6}} & \frac{1}{\sqrt{3}} & \frac{1}{\sqrt{2}}
\end{pmatrix}P,
\label{Vtri}
\end{equation}
where $P$ is a diagonal phase matrix. This form is often called in the
literature the tri/bi-maximal mixing~\cite{tribi}\cite{tribi2} with
the prediction of mixing angles 
$\,\theta_{12}=\sin^{-1}(1/\sqrt{3})\simeq 35.3^\circ$, 
$|\theta_{23}|=45^\circ$, and $\theta_{13}=0^\circ$, which are in
excellent agreement with the latest neutrino experimental
data~\cite{analysis}. Also, CP is conserved in neutrino oscillations.
A remarkable point is that the generation mixing is not affected by
any detail of mass spectrum and model parameters. Furthermore it is
also independent of boundary conditions; the tri/bi-maximal mixing
follows for almost all types of flavor twisting which are allowed by
the consistency relation $X_ZX_T=X_T^{-1}X_Z$, while the mass
eigenvalues are full of variety. Therefore the mixing 
matrix (\ref{Vtri}) is a robust prediction of the presented
mechanism. The Scherk-Schwarz theory of neutrino flavor provides the
observed mixing pattern quite naturally.

\medskip
It is enlightening here to discuss how the tri/bi-maximal mixing is
derived in our scheme of flavor symmetry breaking. In the basis where
the weak current is flavor blind, the light neutrino mass 
matrix ${\cal M}_l$ should take the form
\begin{equation}
  {\cal M}_l \;=\; V^*
  \begin{pmatrix}
    m_1 & & \\[1mm] & \!m_2 & \\[1mm] & & \!\!m_3
  \end{pmatrix} V^\dagger \;=\;
  \frac{m_1'}{6}\,E+\frac{m_2'}{3}\,D+\frac{m_3'}{2}\,F,
\end{equation}
where the mass parameters with primes include Majorana phases. It is
immediately noticed that several integer matrices must be realized and
so highly correlated mass couplings are required to have the
tri/bi-maximal mixing. For this purpose, some explicit models have
been elaborated~\cite{models}. In addition, the mixing 
matrix $V$ seems difficult to be obtained from a viewpoint of flavor
symmetry (breaking). To see this, it is relevant to 
rewrite ${\cal M}_l$ as
\begin{equation}
  {\cal M}_l \;=\; \frac{m_1'+m'_3}{2}\,I+\frac{m_2'-m_1'}{3}\,D
  +\frac{m_1'-m'_3}{2}\,X_4.
  \label{Ml}
\end{equation}
The first and second terms preserve the $S_3$ flavor symmetry, while
the last term breaks $S_3$ down to $S_2$ which acts as an exchange of
the second and third generations. To have the last term of symmetry
breaking is a key ingredient for the tri/bi-maximal mixing. There are
two awkward implications of this desired form of ${\cal M}_l$. The
first is the magnitude of flavor symmetry breaking: the last term in
(\ref{Ml}) becomes a dominant contribution for any type of allowed
neutrino mass spectrum. It seems unlikely in usual model building that
symmetry breaking parameters are larger than symmetric ones. In our
scheme, the decoupling of many heavy fermions amplifies breaking
effects. Another problem is the pattern of flavor symmetry breaking:
the last term in (\ref{Ml}) is not general for the unbroken
symmetry. The most generic $S_2$-invariant matrix is
\begin{equation}
  \alpha I +\frac{\beta}{3} D \,+
  \begin{pmatrix}
    \gamma & & \\ & & \delta \\ & \delta &
  \end{pmatrix}.
\end{equation}
Eq.~(\ref{Ml}) indicates that a special type of flavor symmetry 
breaking, $\gamma=\delta$, is required for the tri/bi-maximal mixing.
If $\gamma\neq\delta$, the 1-2 mixing angle falls into being outside
the experimental range: a small difference $|\gamma-\delta|$ gives a
big effect on the 1-2 mixing angle (and the other angles are not
modified). This behavior comes from the fact that the prediction of
tri/bi-maximal mixing has no relations with mass eigenvalues. The
experimental bound on 1-2 mixing 
implies $\big|1-\frac{\gamma}{\delta}\big|\lesssim
0.12\,(\Delta m^2_{21}/\Delta m^2_{31})^{\frac{1}{2}\div1}\lesssim
O(0.01)$, depending on the neutrino mass spectrum. The
relation $\gamma=\delta$ must be therefore exact but does not seem to
follow from any symmetry argument. We have discussed that the
necessary form of mass matrix is highly complicated and also
technically unnatural from a viewpoint of flavor symmetry
(breaking). The right flavor-breaking effect is readily available in
our five-dimensional theory of flavor with boundary condition twisting.

\medskip
Most types of boundary conditions are found to lead to the mixing
matrix (\ref{Vtri}). The general discussion above suggests that a
possible deviation~\cite{dev} from the tri/bi-maximal mixing appears
in the case that flavor symmetry is completely broken. Among the
boundary twisting allowed by the consistency relation, there exists
only one type of such exceptional twisting on bulk 
fermions; $(X_Z,X_T)=(X_4,X_2)$ [or physically equivalent 
one $(X_4,X_3)\,$]. Fourier-expanding the bulk fermions, evaluating
mass matrices, and integrating over the extra space, we find the
following Majorana mass matrix for three-generation light neutrinos:
\begin{eqnarray}
  {\cal M}_l &=&
  \frac{\frac{|a+b|}{a+b}(p+r)^2}{6\tanh(|a+b|\pi R)}\,D
  +\frac{\frac{|a|}{a}p^2\sinh(2|a|\pi R)}{12\cosh(2|a|\pi R)+6}\,E
  \nonumber\\[2mm]
  && \qquad
  -\frac{\frac{a}{|a|}\bar p^2\sinh(2|a|\pi R)}{4\cosh(2|a|\pi R)+2}\,F
  +\frac{p\bar p}{4\cosh(2|a|\pi R)+2}\,G,
\end{eqnarray}
where $m^c=\bar pI+\frac{\bar r}{3}D$, and we have defined
\begin{equation}
  G \;=\; 
  \begin{pmatrix}
   & 1 & -1 \\[1mm] 1 & -1 & \\[1mm] -1 & & 1
 \end{pmatrix}.
\end{equation}
The last $G$ term does not preserve the previous $S_2$ symmetry
(the $2\leftrightarrow3$ exchange), and the tri/bi-maximal mixing is
expected to be modified. It is noticed that $G$ has 
another $S_2$ invariance ($1\leftrightarrow2$ 
or $1\leftrightarrow3$ exchange) if suitably combined 
with $D$, $E$, and $F$. This indicates that the flavor symmetry has
two breaking sources, i.e.\ the non-trivial local twisting 
at $x^5=0\,$ ($X_Z\neq I$) and $x^5=\pi R\,$ ($X_{Z'}\!\neq\! I,X_Z$),
and flavor symmetry is entirely broken only globally. In an
approximation that $S_2$ breaking (the $G$ term) is sub-leading, the
neutrino mixing matrix deviates from the tri/bi-maximal form as
\begin{equation}
  V' \;=\; V
  \begin{pmatrix}
    1 & & \epsilon \\[1mm] & \!1 & \\[1mm] -\bar\epsilon & & 1
  \end{pmatrix}P' \;=\;
  \begin{pmatrix}
    \frac{2}{\sqrt{6}} & \;\frac{1}{\sqrt{3}}\; & 
    \frac{2\epsilon}{\sqrt{6}} \\[2mm]
    \frac{-1}{\sqrt{6}}+\frac{\bar\epsilon}{\sqrt{2}} & 
    \frac{1}{\sqrt{3}} & 
    \frac{-1}{\sqrt{2}}-\frac{\epsilon}{\sqrt{6}} \\[2mm]
    \frac{-1}{\sqrt{6}}-\frac{\bar\epsilon}{\sqrt{2}} & 
    \frac{1}{\sqrt{3}} & \frac{1}{\sqrt{2}}-\frac{\epsilon}{\sqrt{6}}
  \end{pmatrix}P',
\end{equation}
with a small parameter $\epsilon=
\sqrt{3}\omega/2(1+|\omega|^2)\sinh(2|a|\pi R)$ where 
$\omega=|a|p/a\bar p$. Unlike the previous case, non-vanishing,
complex-valued 1-3 matrix element is generated. This implies the
mechanism simultaneously generates CP violation as well as flavor
symmetry breaking. It should be mentioned that the 
parameter $\epsilon$ cannot be made real by field redefinition; even
if lepton doublets are freely rotated, a combination $\omega$ contains
a unremovable physical phase degree of freedom.

The corrections to mixing angles depend on the neutrino mass
spectrum. We here examine the NH case, and the IH is obtained by
simply exchanging $m_1\leftrightarrow m_3$ (and $a\leftrightarrow a^*$, 
$p\leftrightarrow\bar p$). The mass eigenvalues with NH become
\begin{eqnarray}
  |m_1| &\simeq& \frac{\,|p|^2\big[\sinh^2(2|a|\pi R)
    +\frac{3}{2}\big]^{\frac{1}{2}}}{2\cosh(2|a|\pi R)+1}, \\[.5mm]
  |m_2| &=& \frac{|p+r|^2}{2\tanh(|a+b|\pi R)}, \\[2mm]
  |m_3| &=& \frac{|\bar p|^2\sinh(2|a|\pi R)}{2\cosh(2|a|\pi R)+1},
\end{eqnarray}
with $|p|\ll |\bar p|\tanh^2(2|a|\pi R)$. The smallest 
eigenvalue $m_1$ is found to receive an $S_2$-breaking effect in the
case of small Majorana mass ($|a|R\ll1$). The neutrino mixing angles
are controlled by only one parameter $\epsilon$ and have rather
definite predictions. First notice that, even if the couplings in the
theory are exactly flavor blind ($b,r\to0$), one can obtain sizable
generation mixing of low-energy neutrinos as well as realistic
non-degenerate mass spectrum. The 1-2 mixing is still given by that of
the tri/bi-maximal mixing, similar to all the other boundary
conditions ($\theta_{12}\simeq 35.3^\circ$). Therefore this is a
robust prediction of our scheme for flavor symmetry breaking. The 1-3
mixing angle is 
small; $\theta_{13}=\big|\frac{2}{\sqrt{6}}\epsilon\big|$. For a small
bulk Majorana mass $|a|R\ll 1$ and with flavor-blind couplings, a
predictive relation among the observables is 
derived; $\sin^2\theta_{13}=\frac{1}{6}
(\Delta m^2_{21}/\Delta m^2_{31})^{1/2}$. For a larger Majorana mass,
the 1-3 mixing is suppressed by $\sinh(2|a|\pi R)$ and would not be in
detectable range in future experiments. The low-energy leptonic CP
violation is determined by the complex phase 
of $\epsilon$ parameter ($\phi_\epsilon$), which is the only one
unremovable phase in the lepton sector as stated above, and may also
be concerned with other CP non-conserving phenomena such as baryon
asymmetry in the universe. Finally, the 2-3 neutrino mixing is given
by $\tan\theta_{23}=|(1+\epsilon/\sqrt{3})/(1-\epsilon/\sqrt{3})|$.
Interestingly, the deviation from the maximal mixing is related to the
Dirac CP phase as $\cos2\theta_{23}=
-\sqrt{2}\sin\theta_{13}\cos\phi_\epsilon$. For the minimal CP
violation, that means $6^\circ<|\theta_{23}-45^\circ|<8^\circ$ for NH,
and is comparable with the current experimental bound. On the other
hand, the maximal CP violation implies the maximal 2-3 mixing. The
effective Majorana mass for neutrino-less double beta decay 
is $|\langle m_{ee}\rangle|\simeq|m_2/3+m_3\sin^2\theta_{13}|
\lesssim10^{-2}\,\text{eV}$ and would be accessible in future experiments.

\medskip
The whole charged-lepton sector resides in usual 
four-dimensional (boundary) world. We here present an example of
charged-lepton mass matrix with which the observed lepton mixing is
dominated by the tri/bi-maximal mixing. Similar or more reasonable
models might be available in a variety of scenarios which have been
discussed in four or higher-dimensional theory (though, in more than
five dimension, some regularization~\cite{brane} is needed to
integrate out heavy modes). Our illustrative example introduces a
flavor-triplet neutral scalar $\Phi$ in addition to the right-handed
charged leptons and Higgs doublet which are flavor singlets. The
charged-lepton masses come from the operator involving $\Phi$ with
relevant symmetry imposed. If flavor-breaking expectation 
values $\langle\Phi\rangle=(f_1,f_2,f_3)$ with $f_1\ll f_2\ll f_3$ and
no flavor structure (possibly as well as the neutrino sector) are
utilized, the charged-lepton mass matrix becomes
\begin{equation}
  {\cal M}_e \;=\;
  \begin{pmatrix}
    x_1 & x_2 & x_3 \\[1mm] y_1 & y_2 & y_3 \\[1mm] z_1 & z_2 & z_3
  \end{pmatrix}
\end{equation}
with $x_i\ll y_j\ll z_k$. This induces tiny corrections to
the exact tri/bi-maximal mixing; $\delta\theta_{12}\sim
\frac{m_e}{m_\mu}$, $\delta\theta_{23}\sim\frac{m_\mu}{m_\tau}$, 
and $\delta\theta_{13}\ll1$. An application to the quark sector is 
straightforward. A similar and realistic hierarchy is obtained if the
triplet $\Phi$ accompanies with the right-handed up quarks and the
quark doublets (or only with the right-handed down quarks), which may
also be realized by symmetry argument.

\medskip
To summarize, we have presented a framework for breaking flavor
symmetry with Scherk-Schwarz twist and for communicating it to
low-energy effective theory by integrating out twisted variables. As
an application, the $S_3$ flavor symmetry in the neutrino sector has
been discussed and found to provide rich and robust phenomenology such
as the tri/bi-maximal generation mixing. Other possibilities including
quarks, grand unification, and cosmological study will be investigated
in a separate article.

\bigskip\bigskip
\subsection*{Acknowledgments}
\noindent
This work is supported by Alexander von Humboldt foundation,
scientific grant from the Ministry of Education, Science, Sports, and
Culture of Japan (No.~17740150), and grant-in-aid for the scientific
research on priority area (\#441) "Progress in elementary particle
physics of the 21st century through discoveries of Higgs boson and
supersymmetry" (No.~16081209).


\end{document}